\documentclass[preprint,floatfix,aps,prb,showpacs]{revtex4}  
\usepackage{graphicx}  
  
\newcommand{\beq}{\begin{equation}}  
\newcommand{\eeq}{\end{equation}}  
\newcommand{\beqa}{\begin{eqnarray}}  
\newcommand{\eeqa}{\end{eqnarray}}

\begin{document}  
  
\title{Theory of spin, electronic and transport properties of  
the lateral triple quantum dot molecule in a magnetic field}  
  
\author{F. Delgado}  
\affiliation{Quantum Theory Group, Institute for Microstructural Sciences,  
        National Research Council, Ottawa, Ontario, Canada K1A 0R6}  
  
\author{Y.-P. Shim}  
\affiliation{Quantum Theory Group, Institute for Microstructural Sciences,  
        National Research Council, Ottawa, Ontario, Canada K1A 0R6}  
  
\author{M. Korkusinski}  
\affiliation{Quantum Theory Group, Institute for Microstructural Sciences,  
        National Research Council, Ottawa, Ontario, Canada K1A 0R6}

\author{P. Hawrylak}  
\affiliation{Quantum Theory Group, Institute for Microstructural Sciences,  
        National Research Council, Ottawa, Ontario, Canada K1A 0R6}

\begin{abstract}  
We present a theory of spin, electronic and transport properties  
of a few-electron lateral triangular triple quantum dot molecule in  
a magnetic field.  
Our theory is based on a generalization of a Hubbard model and the 
Linear Combination  of Harmonic Orbitals combined with Configuration 
Interaction method (LCHO-CI) for arbitrary magnetic fields.  
The few-particle spectra obtained as a function of the magnetic field 
exhibit Aharonov-Bohm oscillations. 
As a result, by changing the magnetic field it is possible to engineer 
the degeneracies of single-particle levels, and thus control the total 
spin of the many-electron system.   
For the triple dot with two and four electrons we find oscillations of 
total spin due to the singlet-triplet transitions occurring 
periodically in the magnetic field.  
In the three-electron system we find a transition from  
a magnetically frustrated to the spin-polarized state.  
We discuss the impact of these phase transitions on the addition   
spectrum and the spin blockade of the lateral triple quantum dot 
molecule.  
\end{abstract}  
  
  
\maketitle

\section{Introduction}  
 
There is currently interest in developing the ability to control and 
manipulate the total spin of individually localized interacting 
electrons as a prerequisite for solid-state nanospintronic and quantum 
information  
applications.\cite{awschalom_loss_snt2002,brum_hawrylak_sm1997,loss_divincenzo_pra1998,divincenzo_bacon_nature2000,sachrajda_hawrylak_book2003}  
Precise control over the number and spatial location of carriers  
can be achieved by confining them in lateral gated quantum dot devices,  
of which the  
single,\cite{ciorga_sachrajda_prb2000,tarucha_austing_prl1996}  
double,\cite{holleitner_blick_science2002,pioro_abolfath_prb2005,  
koppens_folk_science2005,petta_johnson_science2005,hatano_stopa_science2005}  
and triple\cite{vidan_westervelt_apl2004,vidan_westervelt_jsupercond2005,gaudreau_studenikin_prl2006,ihn_sigrist_njphys2007}  
quantum dots have already been demonstrated.  
In particular, Gaudreau {\em et al.}\cite{gaudreau_studenikin_prl2006,korkusinski_gimenez_prb2007}  
reported a controlled charging of a lateral triple quantum dot (TQD) 
molecule with $N=1-6$ electrons, with the ability to control the 
population of each dot independently. 
Preliminary experiments on quantum dot molecules in external magnetic
field by Gaudreau {\em et al.}\cite{gaudreau_sachrajda_icps2006} and by   
Ihn {\em et al.}\cite{ihn_sigrist_njphys2007} showed signatures 
of Aharonov-Bohm (AB) oscillations, indicating 
coherent coupling between the constituent dots. 
In this work we present a theory of the magnetic field effect on the
electronic, spin, and transport properties of an isolated triple
quantum dot molecule with controlled  number of electrons $N=1-6$.  
 
%%%%%%%%%%%%%% state of knowledge i.e. previous theoretical work on tqd in B %%%%%% 
 
Previous theoretical descriptions of isolated lateral multi-quantum
dot devices in a magnetic field focused on quantum dot molecules with
one electron per dot using Hubbard, exact numerical diagonalization,
and spin Heisenberg
model.\cite{scarola_dassarma_pra2005,scarola_park_prl2004}
They showed magnetic field induced corrections to the Heisenberg model
due to chiral spin interactions. 
Furthermore, for three dots in a triangular structure, one electron each, 
they established magnetic field-induced transitions from 
a lowest-energy spin doublet with total spin $S=1/2$
(Ref.~\onlinecite{hawrylak_korkusinski_ssc2005}) 
to $S=3/2$ spin polarized
state.\cite{scarola_park_prl2004,scarola_dassarma_pra2005}   
Spin transitions in isolated lateral multi-quantum dot devices with
large electrons numbers have also been studied using spin density
functional theory by Stopa {\em et al.}\cite{stopa_vidan_physe2006}
There has also been significant interest in triple quantum dots in
triangular configuration connected to the leads. 
Using the Hubbard model the effects of the magnetic field on 
the conductance through an empty and singly occupied triple dot were
studied, with focus on the interplay between the Kondo physics,
symmetries, and the AB   
oscillations.\cite{kuzmenko_kikoin_prl2002,kuzmenko_kikoin_prl2006,kikoin_avishai_condmatt2006,jiang_sun_jphysc2007,emary_condmatt2007}
%%%%%%%%%%% summary of what our contribution is  &&&&&&&&&&&&&&&&&&&&&&&&&&  
 
The aim of this work is to study the magnetic field dependence of the 
electronic properties of the lowest electronic shell of a triangular
triple quantum dot molecule filled with $N=1-6$ electrons, extending
in this way our previous work\cite{korkusinski_gimenez_prb2007}  
to finite magnetic fields. 
This is accomplished by both the analysis of the Hubbard model and by
the development of a new computational tool. 
The new microscopic tool combines (i) a calculation of  single
particle states as a linear combination of harmonic orbitals (LCHO)  
localized on each dot,  with a proper gauge transformation allowing
for convergent results as a function of the ratio of the magnetic
length to the inter-dot separation, with (ii)
configuration-interaction approach (CI) to the many-electron problem. 
These techniques have allowed us to analyze the spin and electronic
properties as a function of the magnetic field and the number $N$ of
confined electrons ($N=1$ up to $6$). 
We derive the magnetic-field evolution of the one-electron spectrum 
and show the existence of degeneracies at 
multiples of half flux quanta threading the area of the TQD,  
in agreement with Ref.~\onlinecite{kuzmenko_kikoin_prl2002}. 
The magnetic field-engineered degeneracies of single-particle levels,
combined with electron-electron exchange and correlations,   
allow for the control of the total spin of the many-electron system. 
For example, we show total spin oscillations due to the
singlet-triplet transitions occurring periodically in the magnetic
field for two and four electron molecules.  
In the three-electron system we find the magnetic field-induced
transition from a magnetically frustrated to the spin-polarized state,
in agreement with
Refs.~\onlinecite{scarola_park_prl2004,scarola_dassarma_pra2005}.  
We discuss the impact of these spin transitions on the addition   
spectrum as measured using charge spectroscopy, and predict the
appearance of spin blockade in the transport through TQD molecule.   
  
%%%%%%%%%%%%% organization of paper %%%%%%%%%%%%%%%%%%%%%%%%%%  
  
The paper is organized as follows.  
In Sec.~\ref{secmodel} we present details of the Hubbard and LCHO-CI 
approaches.   
In Sec.~\ref{sec1to6} we calculate the electronic structure of the  
triple dot filled with $N=1$ to $6$ electrons as a function of the  
magnetic field.  
The discussion of the charging diagram and addition amplitudes is  
presented in Sec.~\ref{seccharging}.  
The paper is summarized in Sec.~\ref{summary}.  
  
%%%%%%%%%%%%%%%%%%%%%%%%%%%%%%%%%%%%%%%%%%%%%%%%%%%%%%%%%%%%%%%%%%%%%  
\section{The model \label{secmodel}}  
%%%%%%%%%%%%%%%%%%%%%%%%%%%%%%%%%%%%%%%%%%%%%%%%%%%%%%%%%%%%%%%%%%%%%  

A schematic picture of the TQD studied in this 
work is shown in  Fig.~\ref{fig0}(a).  
This system is an approximation of the lateral gated TQD 
device, in which the three potential minima are created  
electrostatically by metallic gates.  
Such device has been studied theoretically in  
Refs.~\onlinecite{hawrylak_korkusinski_ssc2005,korkusinski_gimenez_prb2007}  
and is related to the system demonstrated experimentally by Gaudreau {\em et  
  al.}\cite{gaudreau_studenikin_prl2006}  
Figure~\ref{fig0}(b) shows the TQD electrostatic potential generated
by a model arrangement of gates enclosing the area of the device (not
shown) together with additional gates (shown as white regions) used to
establish the potential barriers between the dots.
By selective tuning of the voltages it is possible to bring the 
three dots into resonance, i.e., match the energies of the lowest 
single-particle orbital of each potential minimum.  
The resulting TQD molecule can be then filled 
controllably with $N$ electrons, starting at $N=1$, in the presence of 
a magnetic field ${\bf B}=[0,0,B]$ applied in the direction 
perpendicular to the plane of the system.   
 
\subsection{Hubbard model}  
  
We have shown previously\cite{korkusinski_gimenez_prb2007} that the  
electronic properties of the molecule with few confined electrons  
($N=1$ to $6$)  can be understood in the frame of the Hubbard model.  
Assuming one orbital with energy $E_{i,\sigma}$ in each dot, the  
Hubbard Hamiltonian can be written as  
\begin{equation}  
  \hat{H}_H = \sum_{\sigma,i=1}^3 E_{i,\sigma} c_{i\sigma}^\dag c_{i\sigma}  
+ \sum\limits_{\sigma,i,j=1,i\neq j}^{3}  
             t_{ij}(B)c_{i\sigma}^\dag c_{j\sigma}  
+ U\sum\limits_{i=1}^{3} n_{i\downarrow} n_{i\uparrow}  
+ {1\over{2}} V \sum\limits_{i,j=1,i\neq j}^{3} \varrho_i \varrho_j,  
\label{hubhamil}  
\end{equation}  
where the operators $c_{i\sigma}$ ($c_{i\sigma}^\dag$) annihilate 
(create) an electron  with spin $\sigma=\pm 1/2$ in dot $i$. 
Further, $n_{i\sigma} = c_{i\sigma}^\dag c_{i\sigma}$  
and  
$\varrho_i = n_{i\downarrow} + n_{i\uparrow}$  
are, respectively, the spin and charge density on the $i$th dot.  
In Eq. (\ref{hubhamil}), $t_{ij}(B)$ is the  matrix element describing 
tunneling between dots $i$ and $j$, $U$ is the onsite Coulomb 
repulsion, and $V$ is the direct repulsion of electrons occupying 
neighboring dots. 
  
In the Hamiltonian (\ref{hubhamil}) the magnetic field $B$ is 
accounted for in two terms. 
First, it introduces the Zeeman splitting in the onsite energies 
$E_{i,\sigma} = E_i+ g^* \mu B \sigma$, with $g^*$ being the 
effective Land\'e 
factor and $\mu_B$ being the Bohr magneton.   
Second, it renormalizes the single-particle tunneling elements $t_{ij}$ 
by Peierls phase factors,\cite{peierls_zphys1933,luttinger_pr1951}  
such that $t_{ij}(B)=t_{ij}e^{2\pi i \phi_{ij}}$.  
For the three quantum dots located in the corners of an  
equilateral triangle we have $\phi_{12}=\phi_{23}=\phi_{31}=-\phi/3$. 
Here, $\phi=3\sqrt{3} e B R^2 / 8\pi\hbar c$ is the number of magnetic 
flux quanta threading the system, with $e$ being the electron charge, 
$c$ - the speed of light, $\hbar$ - the Planck's constant, and  
$R$ - the distance from the center of the triangle to each dot, 
identified in Fig.~\ref{fig0}(a).   
  
With one spin-degenerate orbital per dot we can fill the TQD with up 
to $N=6$ electrons.  
To find the eigenenergies and eigenstates of $N$ electrons  
we use the configuration interaction approach (CI), in which  
we create all possible configurations of $N$ electrons on the localized  
orbitals, write the Hamiltonian $\hat{H}_{H}$ in a matrix form in  
this basis, and diagonalize the resulting matrix 
numerically.\cite{korkusinski_gimenez_prb2007}

\subsection{LCHO-CI method}  
 
We compare the results of the Hubbard model with a microscopic approach 
to the calculation of the electronic properties of a TQD starting from 
a confining potential, which we outline in this section. 
We start by expressing  all energies in units of the effective  
Rydberg ${\cal R}=m_e^* e^4 / 2 \varepsilon^2 \hbar^2$, and lengths in  
the effective Bohr radius, $a_B = \varepsilon \hbar^2 / m_e^* e^2$,  
where $m^*$ is the electron effective mass and $\varepsilon$ is the 
dielectric constant of the material.   
With GaAs parameters, $m_e^*=0.067$ $m_0$ and $\varepsilon=12.4$, we  
have ${\cal R}= 5.93$ meV and $a_B = 9.79$ nm.  
  
A single electron in the TQD in the presence of an external 
perpendicular magnetic field is described by the Hamiltonian   
\begin{equation} 
\label{eq:H_SP}  
\hat{H}_0 = \left( -i\nabla + \mathbf{A}(\mathbf{r}) \right)^2  
          + \sum_{i=1}^{3} V_i(\mathbf{r}) + V_B(\mathbf{r})  
\end{equation}  
where $\mathbf{A}(\mathbf{r})$ is the effective vector potential,  
$V_i(\mathbf{r})$ is the confining potential of the $i$-th dot,  
and $V_B(\mathbf{r})$ is the potential due to the additional gates, 
which control the potential barriers between dots.  
  
\begin{figure}[h]
  \includegraphics[width=0.8\textwidth]{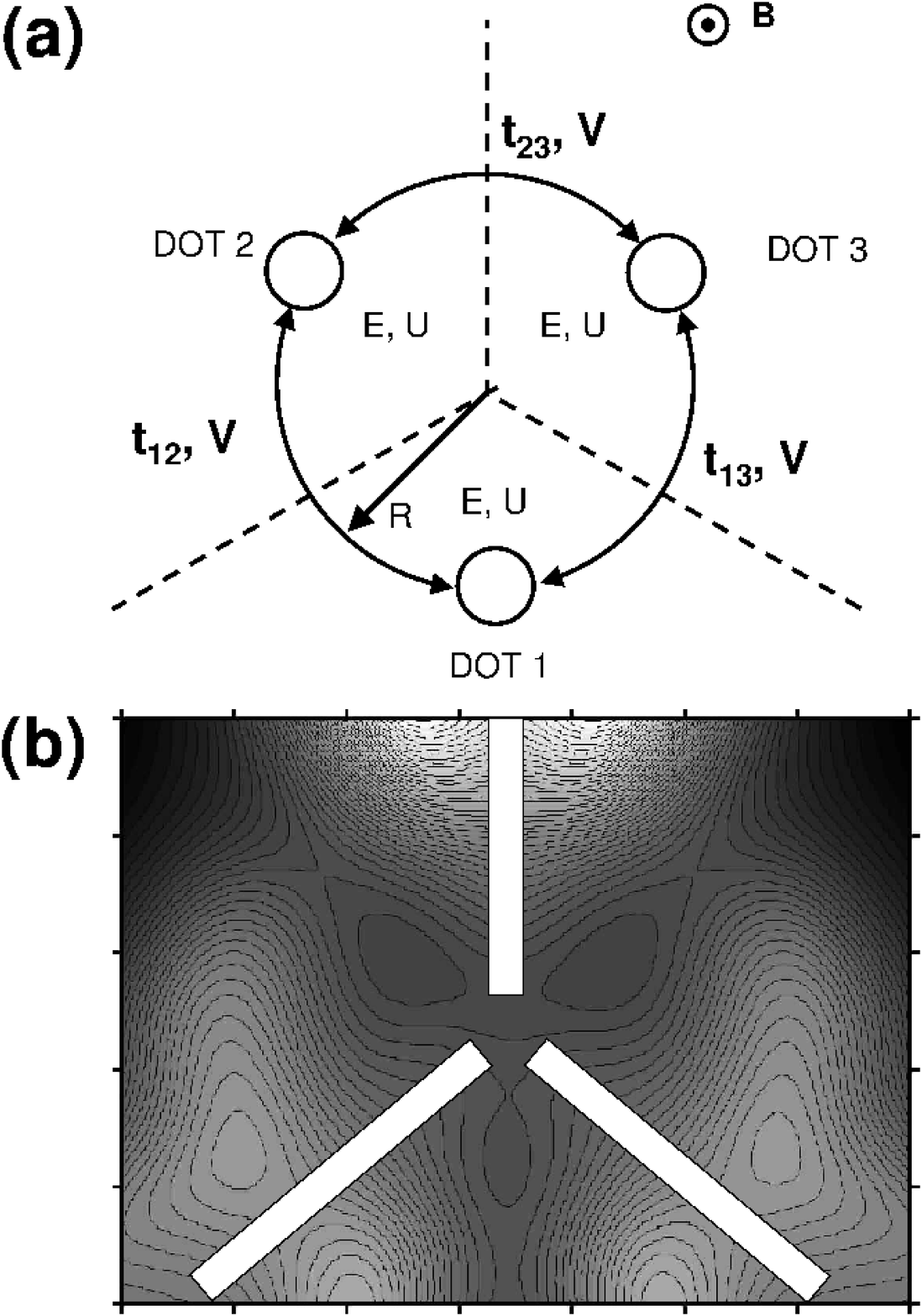}
  \caption{(a) Schematic image of the triple-dot molecule.  
    Symbols denote the Hubbard parameters for the system on resonance.  
    The magnetic field is perpendicular to the plane of the molecule. 
    (b) Contour plot of the potential created in the lateral
    triple-dot device by a typical  layout of the metallic gates. 
    White zones show schematically the gates controlling the
    barriers between dots.}
  \label{fig0}  
\end{figure}

\begin{figure}[h]
  \includegraphics[width=0.8\textwidth]{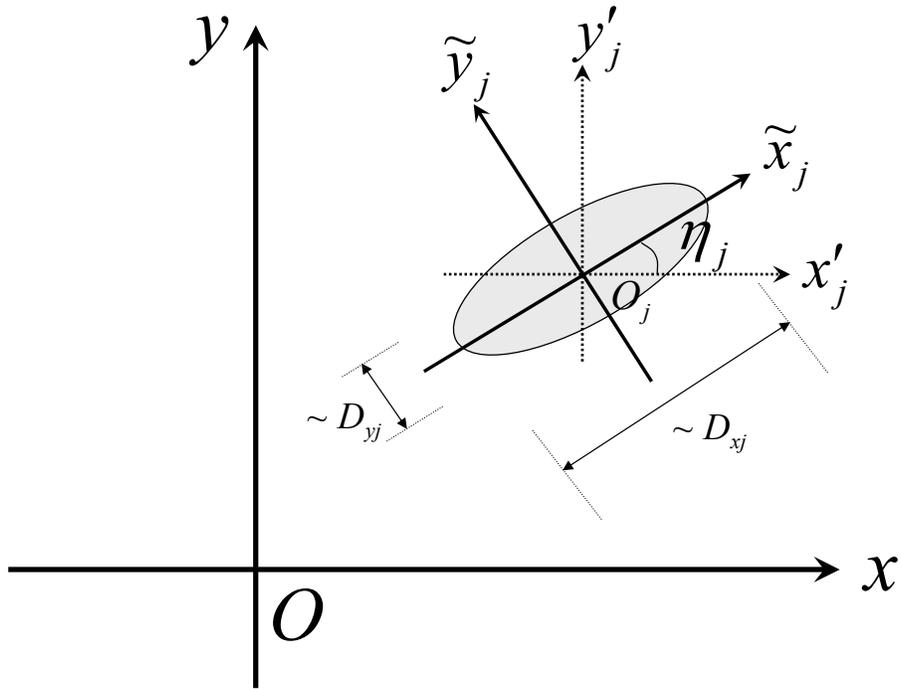}
  \caption{Schematic picture of the potential created by one of the Gaussian  
  barrier centered at $(x_{Bj},y_{Bj})$ and the relation between the  
  original variables $(x,y)$ and the system $(\tilde{x}_j,\tilde{y}_j)$.}  
  \label{fig0_1}  
\end{figure}

We choose the vector potential in the symmetric gauge 
$\mathbf{A}(\mathbf{r})=\Omega_c/4(-y\hat{\mathbf{x}}+x\hat{\mathbf{y}})$ 
centered at the geometric center of the triangle of the 
dots.\cite{scarola_park_prl2004}   
Here the cyclotron energy $\Omega_c=\hbar\omega_c/{\cal R}$ with 
$\omega_c=eB/m_e^*c$.   
The confining potential of each dot is approximated by a Gaussian  
$V_i(\mathbf{r})=-V_{i,0}\exp\left[-\left({\mathbf{r}-\mathbf{r}_i 
      \over d_i}  \right)^2\right]$.   
Further we separate the Gaussian potential into the harmonic and 
anharmonic parts,  
\begin{equation}  
V_i(\mathbf{r})  
= -V_{i,0} + \frac{\Omega_{i,0}^2}{4} \left({\mathbf{r}-\mathbf{r}_i \over d_i}  \right)^2 + \delta V_i,  
= V^{HO}_i + \delta V_i  
\end{equation}  
where $\Omega_{i,0}=2\sqrt{V_{i,0}}/d_i$ is the effective 
characteristic energy of the harmonic confinement.   
In order to tune the height of the tunneling barrier between dots  
independently of the confining potential,  
we introduce Gaussian barrier potentials located between each pair of 
dots.\cite{abolfath_hawrylak_prl2006} 
In the device depicted in Fig.~\ref{fig0}(b) these potentials are 
generated by the gates shown as white regions. 
The coordinate system used to define the Gaussian barriers is 
summarized in Fig.~\ref{fig0_1}.  
Here we assume that the narrow gate is oriented along the axis 
$\tilde{x}$, which means that the geometry in Fig.~\ref{fig0_1} 
applies specifically to the lower left-hand gate of  
Fig.~\ref{fig0}(b). 
The Gaussian barrier can now be defined as  
\begin{equation}  
V_B(\mathbf{r})=\sum_{j=1}^3 V_{B}^{(j)}(\mathbf{r})  
 = V_{B0}^{(j)} \mathrm{exp}\left( - \frac{\tilde{x}_{j}^2}{D_{xj}^2} 
   - \frac{\tilde{y}_{j}^2}{D_{yj}^2} \right),   
\end{equation}  
with the global and local coordinate systems related by 
\begin{eqnarray}  
\tilde{x}_{j} &=& (x-x_{Bj})\cos\eta_j + (y-y_{Bj})\sin\eta_j \nonumber\\  
\tilde{y}_{j} &=&-(x-x_{Bj})\sin\eta_j + (y-y_{Bj})\cos\eta_j.  
\label{coordtransform} 
\end{eqnarray}  
If we  choose  both the global gauge $\mathbf{A}(\mathbf{r})$ and a 
computational basis centered at the origin of a 
TQD,\cite{scarola_park_prl2004,scarola_dassarma_pra2005} we find a  
very poor convergence of results as a function of the size of 
single-particle basis, especially for large interdot distances, when 
each dot should essentially be considered separately, with its own 
vector potential.\cite{abolfath_hawrylak_prl2006}   
To remedy this, we divide the system into three regions, whose 
boundaries are marked in Fig.~\ref{fig0}(a) by dashed lines, and in 
each region define the vector potential in the form  
\begin{equation}  
\mathbf{A}_i = \frac{\Omega_c}{4} \left[ -(y-y_{i})\hat{\mathbf{x}} + 
  (x-x_{i})\hat{\mathbf{y}} \right],   
\end{equation}  
i.e., centered in the potential minimum of the respective dot.  
 
We solve for the eigenenergies and eigenvectors of the Hamiltonian 
(\ref{eq:H_SP}) in the basis composed of harmonic oscillator states 
(HO) of each dot in the magnetic field  
\begin{equation}\label{eq:phi_inm}  
\langle \mathbf{r} | inm \rangle = \phi_{inm}(\mathbf{r})  
= \mathrm{exp}\left[-i\frac{\Omega_c}{4}(-y_i x + x_i 
  y)\right]\phi^{(i)}_{nm}(\mathbf{r}-\mathbf{r}_i), 
\end{equation}  
where $\phi^{(i)}_{nm}(\mathbf{r}-\mathbf{r}_i)$ are the HO orbitals 
of  $i$th dot, satisfying the Schr\"odinger equation 
\begin{equation}  
\left[ \left( -i\nabla + \mathbf{A}_i(\mathbf{r}) \right)^2 + V^{HO}_i 
\right]\phi^{(i)}_{nm}(\mathbf{r}-\mathbf{r}_i)   
=\varepsilon_{inm}^{HO} \phi^{(i)}_{nm}(\mathbf{r}-\mathbf{r}_i). 
\label{eigho}  
\end{equation}  
The energy associated with the HO state $\phi^{(i)}_{nm}$, 
\begin{equation}  
\varepsilon_{inm}^{HO} = -V_{i,0} + \Omega_{i,+} 
\left(n+\frac{1}{2}\right) + \Omega_{i,-} \left(m+\frac{1}{2}\right),   
\end{equation}  
is defined in terms of energies $\Omega_{i,\pm} = \Omega_{i,h} \pm 
\Omega_c /2$, with the hybrid energy 
$\Omega_{i,h} =\sqrt{ \Omega_{i,0}^2 + \Omega_c^2/4}$. 
The eigenfunctions of Eq. (\ref{eigho}) are the Fock-Darwin (FD) 
orbitals, whose explicit form as a function of $z=x+iy$ and 
$\bar{z}=x-iy$ is   
\begin{equation}  
\phi^{(i)}_{nm}(\mathbf{r})  
 = \frac{(-1)^{n_2}}{\sqrt{2\pi l_{i,h}^2}} \sqrt{\frac{n_2!}{n_1!}}  
     \,\mathrm{L}_{n_2}^{n_1-n_2} \left( \frac{z\bar{z}}{2l_{i,h}^2} \right)  
      e^{-\frac{z\bar{z}}{4 l_{i,h}^2}}  
     \left\{  
     \begin{array}{l}  
     \left( \bar{z}/\sqrt{2} l_{i,h} \right)^{m-n} \quad\mathrm{for}\quad m \geq n \\  
     \left( z/\sqrt{2}l_{i,h} \right)^{n-m} \quad\mathrm{for}\quad n \geq m  
     \end{array}  
     \right. 
\end{equation}  
Here $n_1=\mathrm{max}(n,m)$, $n_2=\mathrm{min}(n,m)$ and the hybrid 
length $l_{i,h}=\sqrt{1/\Omega_{i,h}}$.   
Further, $\mathrm{L}_{n}^{k}$ is the generalized Laguerre polynomial 
defined as 
\begin{equation}  
\mathrm{L}^k_n (x) = \sum_{l=0}^{n}  
                     \frac{(-1)^{l}(n+k)!}{(n-l)!(k+l)!l!}  
                     x^l~,  
                     \quad\mathrm{for}\quad k>-1~.  
\end{equation}  
The phase factor $e^{-\frac{i\Omega_c}{4}(-y_i x + x_i y)}$ of the 
basis function $\phi_{inm}$ in Eq.~(\ref{eq:phi_inm}) is due to the 
gauge transformation $\mathbf{A}_i \Rightarrow \mathbf{A}$.  
This additional phase factor depends on the magnetic field and the  
distance of dot $i$ from the origin,  
and leads to the flux-dependent factor renormalizing the  
tunneling matrix elements in the Hubbard model.

Now we can represent the single-electron eigenvalue problem of the TQD 
in matrix form  in a restricted Hilbert space formed by $n_0$ FD 
orbitals from each dot, with dimension $N_{orb}=3n_0$, as:   
\begin{equation}  
\mathbf{H}_{0}\mathbf{a}^{(n)} = \varepsilon_n \mathbf{S} \mathbf{a}^{(n)}  
\label{geneig}  
\end{equation}  
where $\mathbf{H}_{0}$ is the Hamiltonian matrix for $\hat{H}_0$ in 
Eq.~(\ref{eq:H_SP}) and $\mathbf{S}$ is the overlap matrix due to the 
non-orthogonality of the basis and $\mathbf{a}^{(n)}$ is an 
eigenvector corresponding to the eigenvalue $\varepsilon_n$.   
The eigenstates are given by  
\begin{equation}\label{eq:psi_LCHO}  
\psi_n(\mathbf{r})=\sum_{k=1}^{N_{orb}} a^{(n)}_k \phi_k(\mathbf{r})  
\end{equation}  
where the composite index $k=\{inm\}$. 
The Hamiltonian and overlap matrix elements can be obtained  
efficiently if we expand the FD orbitals as linear combinations of  
the zero-field HO orbitals $\phi^{(0)}_{n_x,n_y}(\mathbf{r})$ with 
characteristic, magnetic-field dependent energy $\Omega_h$.  
\begin{equation}\label{eq:phi_nm}  
\phi_{nm}(\mathbf{r}) = \sum_{s=0}^{n+m} A_s^{nm} 
\phi^{(0)}_{n+m-s,s}(\mathbf{r}),   
\end{equation}  
where  
\begin{equation}  
A_s^{nm} = \sqrt{\frac{n!m!(n+m-s)!s!}{2^n 2^m}} (-i)^s   
           \sum_{k=\mathrm{max}(0,s-m)}^{\mathrm{min}(s,n)} 
           \frac{(-1)^k}{k!(n-k)!(s-k)!(m-s+k)!}.   
\end{equation}  
Then the integration needed to obtain the Hamiltonian and overlap 
matrix elements can be separated into $x$- and $y$-dependent parts and  
each integral can be carried out analytically. 
For the barrier potential $V_B(\mathbf{r})$ such a separation is 
complicated by the appearance of an $xy$ term in the exponent.  
This term can be eliminated by a transformation to the local 
coordinate system defined in Eq.(\ref{coordtransform}), after which 
the integrals can be carried out analytically. 
  
The generalized eigenvalue problem formulated in Eq.(\ref{geneig}) 
can be cast into a standard eigenvalue problem   
\begin{equation}  
\mathbf{H}' \mathbf{b}^{(n)} = \varepsilon_n \mathbf{b}^{(n)},  
\end{equation}  
where $\mathbf{H}'=(\sqrt{\mathbf{S}})^{-1} \mathbf{H}_{0} 
(\sqrt{\mathbf{S}})^{-1}$   
and $\mathbf{b}^{(n)}=\sqrt{\mathbf{S}} \mathbf{a}^{(n)}$.  
The matrix $\sqrt{\mathbf{S}}$ is found by solving the eigenvalue 
problem  $\mathbf{S}\mathbf{V}_S = \mathbf{V}_S \mathbf{E}_S$.  
Here $\mathbf{V}_S$ is the matrix of eigenvectors and $\mathbf{E}_S$ 
is the diagonal matrix with eigenvalues.   
Then $\sqrt{\mathbf{S}}$ is obtained by 
$\sqrt{\mathbf{S}}=\mathbf{V}_S^{\dag} \mathbf{E}_S^{1/2} 
\mathbf{V_S}$.   
The off-diagonal elements of the effective Hamiltonian $\mathbf{H}'$  
correspond to the tunneling elements in the Hubbard model.  
To see how the gauge transformation automatically takes care of the 
phase change of the tunneling elements let us consider a resonant TQD 
system where all three confining potentials are identical.   
Since we will consider only $s$ orbitals from each dot (i.e., 
$n=m=0$), we shall use the simplified notation $|i00\rangle \equiv 
|i\rangle$.   
Then the off-diagonal matrix element of the Hamiltonian $\mathbf{H}_0$ 
are 
\begin{equation}\label{eq:H0_ij}  
\langle i | \hat{H}_0 | j \rangle  
 = \Omega_0 \langle i|j \rangle  
  +\langle i | - \frac{\Omega_0^2}{4} (\mathbf{r}-\mathbf{r}_j)^2 | j \rangle  
  +\langle i | \sum_{k=1}^3 V_k(\mathbf{r}) | j\rangle~.  
\end{equation}  
The overlap matrix element takes the form 
\begin{equation}  
\langle i|j \rangle  
= \exp \left[ \frac{i\Omega_c}{4} 
  \hat{\mathbf{z}}\cdot(\mathbf{r}_i\times\mathbf{r}_j)   
             -\frac{(\mathbf{r}_i-\mathbf{r}_j)^2}{8} \left( 
               \frac{1}{l_h^2}+\frac{l_h^2 \Omega_c^2}{4} \right)   
           \right]  
\end{equation}  
and the second term in Eq.~(\ref{eq:H0_ij}) is  
\begin{equation}  
\langle i | - \frac{\Omega_0}{4} (\mathbf{r}-\mathbf{r}_j)^2 | j \rangle  
= -\frac{\Omega_0^2 l_h^2}{4} \langle i|j \rangle  
   \left[ 2+\frac{(\mathbf{r}_i-\mathbf{r}_j)^2}{4}  
            \left( \frac{1}{l_h^2}-\frac{l_h^2 \Omega_c^2}{4} \right)  
   \right].  
\end{equation}  
If we neglect the three-center integrals $\langle i | V_k(\mathbf{r}) | 
j \rangle$ for $k\neq i,j$, the last term in Eq.~(\ref{eq:H0_ij}) is 
obtained in the form  
\begin{equation}  
\langle i | V_i(\mathbf{r})+V_j(\mathbf{r}) | j \rangle  
= \frac{-2V_0 d^2}{2l_h^2+d^2}  
  \exp\left[ \frac{i\Omega_c}{4} 
    \hat{\mathbf{z}}\cdot(\mathbf{r}_i\times\mathbf{r}_j)   
    -\frac{(\mathbf{r}_i-\mathbf{r}_j)^2}{2(2l_h^2+d^2)}  
    \left( 1+\frac{d^2}{4l_h^2}+\frac{\Omega_c^2 l_h^2 d^2}{16} \right)  
      \right].  
\end{equation}  
The common overall phase $\frac{\Omega_c}{4} 
\hat{\mathbf{z}}\cdot(\mathbf{r}_i\times\mathbf{r}_j)$   
is proportional to the magnetic field and the area of the parallelogram  
formed by the vectors $\mathbf{r}_i$ and $\mathbf{r}_j$.  
Now Eq.~(\ref{eq:H0_ij}) becomes  
\begin{equation}  
\langle i | \hat{H}_0 | j \rangle  
= -  A_{ij} \exp \left[ \frac{i\Omega_c}{4} 
  \hat{\mathbf{z}}\cdot(\mathbf{r}_i\times\mathbf{r}_j) \right], 
\end{equation}  
where the amplitude $A_{ij}$ has complicated dependence on the magnetic field  
but is generally positive and decreases exponentially as the magnetic 
field increases.   
This exponential decrease is due to the suppression of the overlap of 
orbitals from different dots, resulting from the decrease of the 
effective radius of the wave function with the increasing magnetic 
field. 
The off-diagonal element of the effective Hamiltonian $\mathbf{H}'$ 
differs from that in Eq.~(\ref{eq:H0_ij}) due to the existence of the 
overlap matrix $\mathbf{S}$, but the behavior of the phase and the 
amplitude is the same.   
Thus the tunneling parameter in the Hubbard model in the presence of 
magnetic field acquires a field-dependent phase proportional to the 
flux, and amplitude which decays exponentially with the flux.  
  
The eigenstates of the above single-electron problem are linear 
combinations of the  harmonic oscillator orbitals (LCHO).   
We use these LCHO extended molecular orbitals to solve the 
many-electron problem of the TQD system.   
The Hamiltonian of this system is  
\begin{equation}  
\hat{H}  
 = \sum_{i\sigma}( \varepsilon_i + \varepsilon^Z_{\sigma} ) 
 c_{i\sigma}^{\dag}c_{i\sigma}   
  +\frac{1}{2} \sum_{ijkl\sigma\sigma'}\langle ij|v_c|kl\rangle 
  c_{i\sigma}^{\dag}c_{j\sigma'}^{\dag}c_{k\sigma'}c_{l\sigma}   
\end{equation}  
where $i$, $j$, $k$, $l$ enumerate the LCHO orbitals and   
$\sigma$,$\sigma'$ are spin indices.  
The operators $c^{\dag}_{i\sigma}$ $(c_{i\sigma})$ create (annihilate)  
an electron on the spin-orbital $(i,\sigma)$, while  
$\varepsilon^Z_{\sigma}=g^* m_e^*\Omega_c \sigma / 2m_0$ is the 
Zeeman energy.
In the following discussions, the Zeeman energy is accounted for only 
in the sections corresponding to three electrons and the addition
spectra, where it is responsible for the transition between spin
polarized and spin unpolarized ground state. 
In order to make the transition more clear in the corresponding
figures, we have chosen a model value of $g^*=-0.02$ instead of the
usual value corresponding to GaAs ($g^*=-0.44$).  
 
The second term of the above Hamiltonian is scaled by Coulomb 
interaction matrix elements 
\begin{equation}  
\langle ij|v_c|kl\rangle  
= \int d\mathbf{r} \int d\mathbf{r}'  
  \psi_i^*(\mathbf{r}) \psi_j^*(\mathbf{r}')  
  \frac{2}{|\mathbf{r}-\mathbf{r}'|}  
  \psi_k(\mathbf{r}')\psi_l(\mathbf{r}).  
\end{equation}  
Using the Fourier transformation of the Coulomb interaction,  
\begin{eqnarray}  
\langle ij | v_c | kl \rangle  
 &=& \frac{1}{\pi} \int_0^{\infty} dq \int_0^{2\pi} d\theta_q  
     \int d\mathbf{r} \psi_{i}^*(\mathbf{r}) e^{i\mathbf{q}\cdot\mathbf{r}}\psi_{l}(\mathbf{r})  
     \int d\mathbf{r}'\psi_{j}^*(\mathbf{r}')e^{-i\mathbf{q}\cdot\mathbf{r}'}\psi_{k}(\mathbf{r}') \nonumber\\  
 &=& \frac{1}{\pi} \int_0^{\infty} dq \int_0^{2\pi} d\theta_q  
     \langle i | e^{i\mathbf{q}\cdot\mathbf{r}} | l \rangle  
     \langle j | e^{-i\mathbf{q}\cdot\mathbf{r}}| k \rangle~.  
\end{eqnarray}  
The matrix elements of the plane wave $\langle i | 
e^{i\mathbf{q}\cdot\mathbf{r}} | j \rangle$ are evaluated 
analytically using the expansions (\ref{eq:psi_LCHO} - \ref{eq:phi_nm}) 
of the LCHO orbitals in terms of the zero-field HO orbitals. 
The $q$ and $\theta_q$ integrations are then carried out numerically.  
The Coulomb interaction matrix elements can be used to extract the 
interaction parameters in the Hubbard model.   
  
While the LCHO-CI approach is general, we will illustrate it on  the 
TQD molecule with identical quantum dots.  
For a given number of electrons, we consider all possible 
configurations of electrons in the LCHO orbitals, calculate the 
Hamiltonian matrix in this configuration basis and  diagonalize this 
matrix numerically to find the eigenstates and eigenenergies of the 
interacting many-electron TQD.   
In this paper, we consider single-particle basis formed by $s$ orbitals  
from each dot and filling of the lowest electronic shell with $N=1-6$ 
electrons.

%%%%%%%%%%%%%%%%%%%%%%%%%%%%%%%%%%%%%%%%%%%%%%%%%%%%%%%%%%%%%%%%%%%%%  
\section{Magnetic field behavior of the lowest electronic shell}  
  \label{sec1to6}  
%%%%%%%%%%%%%%%%%%%%%%%%%%%%%%%%%%%%%%%%%%%%%%%%%%%%%%%%%%%%%%%%%%%%%  
  
\subsection{Magnetic field dependence of single-electron spectrum}  
  
Let us start our analysis by discussing the single-particle spectrum  
of the triple dot molecule as a function of the magnetic field.  
In the basis of orbitals $\{|1\rangle, |2\rangle, |3\rangle \}$ 
localized on the respective dots, the Hubbard Hamiltonian for a single 
electron in the TQD on resonance takes a matrix form 
\beqa  
\hat{H}_T = \left[  
  \begin{array}{ccc}  
    E & te^{-2\pi i \phi/3} & te^{2\pi i\phi/3}\\  
    te^{2\pi i \phi/3} & E & te^{-2\pi i \phi/3} \\  
    te^{-2\pi i \phi/3} & te^{2\pi i \phi/3} & E\\  
  \end{array}  
\right].  
\label{h1e}  
\eeqa  
The one-electron  Hamiltonian can be diagonalized by performing the 
Fourier transform of the localized basis $|j\rangle$   
into a plane wave basis $|K\rangle$ as  
$|K\rangle = \sum\limits_{j=1}^{3}{ e^{ i K (j-1) } |j\rangle}$
(Ref.~\onlinecite{hawrylak_prl1993}).
The new basis consists of three states, with $K_1=0$, 
$K_2= 2\pi/3$, and $K_3= -2\pi/3$, given by:  
\beqa  
\left\{  
\begin{array}{ll}  
|{K_1}\rangle =\frac{1}{\sqrt{3}}\left(|1\rangle+|2\rangle  
+|3\rangle\right)  
\\  
|{K_2 }\rangle =\frac{1}{\sqrt{3}}\left(|1\rangle  
+e^{i 2\pi/3}|2\rangle+e^{i 4\pi/3}|3\rangle\right)  
\\  
|{ K_3}\rangle =\frac{1}{\sqrt{3}}\left(|1\rangle  
+e^{-i2\pi/3}|2\rangle+e^{-i4\pi/3}|3\rangle\right).  
\end{array}\right.  
\label{eigenvectors}  
\eeqa  
The corresponding  eigenenergies are, respectively: 
$E_1=E-2|t|\cos\left(2\pi\phi/3\right)$,  
$E_2=E-2|t|\cos\left[2\pi(\phi+1)/3\right]$, and  
$E_3=E-2|t|\cos\left[2\pi(\phi-1)/3\right]$. 
At zero magnetic field the three eigenstates  
form a spectrum with a non-degenerate, standing wave (zero effective 
angular momentum) $K=0$ ground state,   
and two degenerate excited states with $K= \pm 2\pi/3$ (effective 
angular momentum $\pm 1$, respectively).   
  
\begin{figure}[h]
  \includegraphics[width=0.8\textwidth]{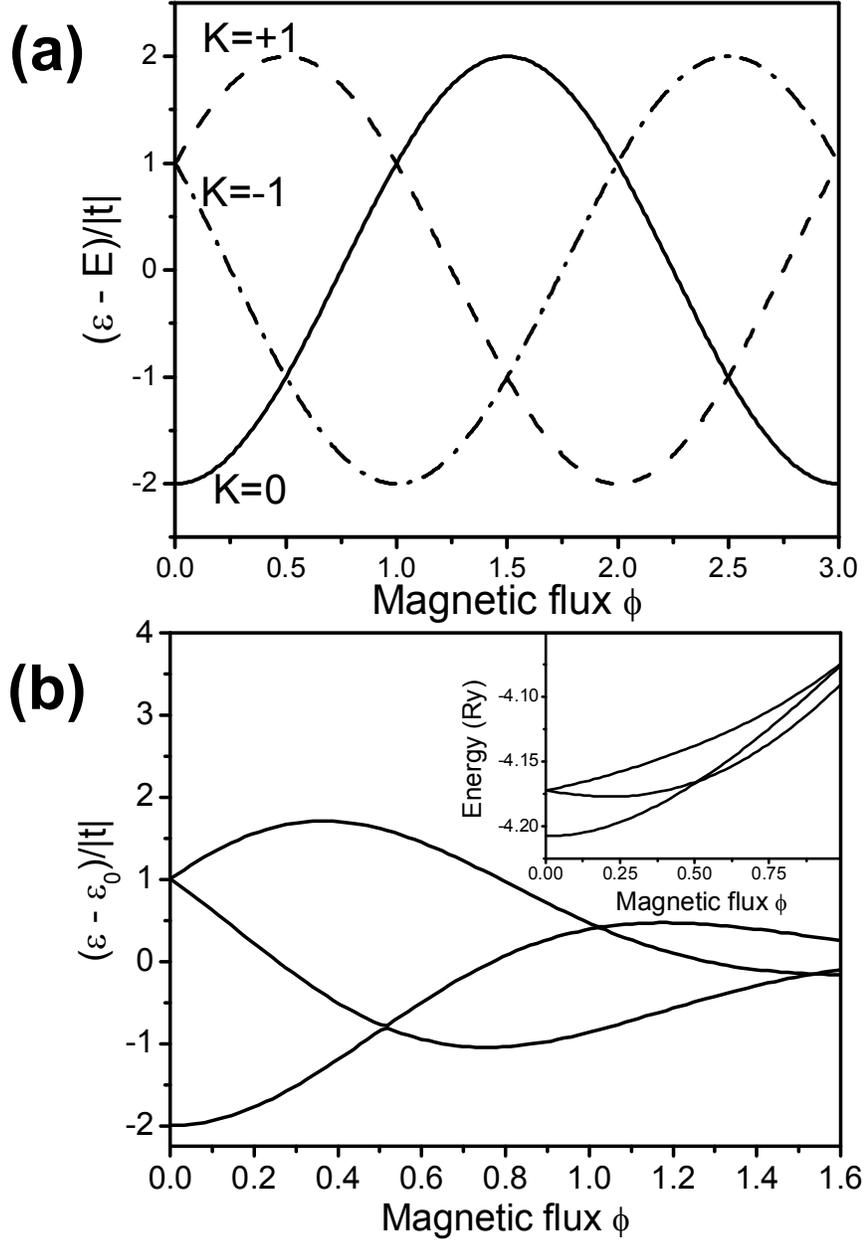}
  \caption{Single-particle energies of the triple dot molecule as a  
    function of the number of magnetic flux quanta calculated within  
    the Hubbard model (a), and with the LCHO approximation (b).  
    In figure (b) the inset corresponds to the actual spectrum while 
    in the main panel the diamagnetic shift has been removed to 
    facilitate the comparison with results from Hubbard model.}  
  \label{fig1}  
\end{figure}

In Fig.~\ref{fig1}(a) we show these energies as a function of the flux  
$\phi$, with different lines corresponding to each effective  
angular momentum.  
The calculations were performed for model parameters $E=0$ and  
$t=-0.0118 {\cal R}$, and in the absence of the Zeeman energy.  
We find that the one-electron energy spectrum is composed of three  
levels, whose energies undergo Aharonov-Bohm oscillations  with period $\Delta \phi=3$ flux  
quanta and amplitude $2|t|$ around the single-dot energy $E$.  
At  $\phi=(2n+1)/2$ with $n=0,1,\dots$ we find a degenerate ground  
state and a nondegenerate excited state of the system.  
On the other hand, for $\phi=n$ the degeneracy is inverted, i.e., the  
ground state  is nondegenerate while the excited state  
is doubly degenerate. The levels correspond to different quantum numbers, and hence  
cross without interaction, leading to degeneracies.  
  
We tested the behavior of the energy spectrum of the Hubbard model  
against the microscopic  LCHO approach.  
We assume the depth of the Gaussian potentials $V_0=5.864$~${\cal R}$,  
their characteristic width $d=2.324$~$a_B$, and the distance between dot  
centers $|\mathbf{r}_i-\mathbf{r}_j|=4.85$~$a_B$  based on fitting to the  
electrostatic confinement produced by a model lateral gated quantum  
dot device.\cite{korkusinski_gimenez_prb2007}  
As discussed in the previous section, the flux-dependent phase factor  
is due to the gauge transformation in LCHO approach.  
%  
%To account properly for the phase change of the tunneling elements in the Hubbard model,  
%we have used and effective radius $R^*$  
%that fits the oscillations in the spectrum of the Hubbard Hamiltonian  
%to those obtained in the LCHO.  
%It was found that the effective radius was approximately $R^* \approx 0.64R$,  
%where $R$ is the radius of the circumference that pass through the center  
%of the three dots.  
%  
In the inset of Fig.~\ref{fig1}(b) we show the single-particle energies  
as a function of the magnetic flux calculated with the  
LCHO method with only one HO orbital per dot.  
The resulting spectrum does exhibit the periodic degeneracies of  
levels.  
It differs, however, from that in Fig.~\ref{fig1}(a) in two aspects.  
First, as a function of the magnetic field all energies undergo a  
diamagnetic shift towards higher energies.  
This shift is, in most part, due to the behavior of single-dot  
energies, which in the LCHO approach are $\varepsilon_{i00}^{HO}=\Omega_h$, and  
therefore increase with the magnetic field.  
In the Hubbard model, on the other hand, we have assumed these  
energies to be constant, irrespective of the number of flux quanta.  
In the main panel of Fig.~\ref{fig1}(b) we have redrawn the LCHO spectrum  
with the diamagnetic shift removed by subtracting the reference energy  
\begin{equation}  
E_0=\langle i00 |  
    \left[ \left( -i\nabla + \mathbf{A}_i(\mathbf{r}) \right)^2  
          + \sum_{j=1}^{3} V_j(\mathbf{r})  
    \right]  
    | i00 \rangle .  
\end{equation}  
The renormalized spectrum  can be directly compared with the energy spectrum  
of the Hubbard model. Both spectra oscillate with increasing magnetic field.  
The second difference between the two spectra involves the amplitude  
of oscillation, which remains constant in the Hubbard approach, but  
decreases in the LCHO treatment.  
This feature can be understood in terms of decrease of the  
magnitude of the effective tunneling parameter $|t|$ with increasing 
magnetic field.   
This can be overcome by reducing the height of tunneling barriers  
between dots using additional gates as discussed  
in Section \ref{secmodel}.  
  
From Fig.~\ref{fig1} it is apparent that by adjusting the magnetic  
field and barrier height we can engineer the degeneracies of the 
single-particle states.   
This property of the triple dot molecule is of key importance when the  
system is being filled with electrons.  
  
\subsection{Two electrons}  
  
Let us start with $N=2$ electrons confined in the triple dot molecule.  
In order to simplify the notation, in the following sections,  
unless the opposite is explicitly stated, we shall denote the complex  
and magnetic field-dependent hopping parameter $t_{ij}(\phi)$  
by $t_{ij}$.  
  
We can classify the two-electron  
states into singlets and triplets according to their total spin.  
Let us start with the triplet subspace, with both electrons spin-down.  
The basis consists of three singly-occupied localized  
configurations:  
$|T_1\rangle=c^\dag_{2\downarrow}c^\dag_{1\downarrow}|0\rangle$,  
$|T_2\rangle=c^\dag_{1\downarrow}c^\dag_{3\downarrow}|0\rangle$,  
$|T_3\rangle=c^\dag_{3\downarrow}c^\dag_{2\downarrow}|0\rangle$.  
Each configuration has the same energy $2E+V$, and each 
pair of  configurations is coupled via the single-particle tunneling 
elements only.  
Therefore, the Hubbard Hamiltonian written in this basis is identical  
to the single electron Hamiltonian, Eq.~(\ref{h1e}),  
except that all off-diagonal tunneling elements acquire a negative 
phase.   
As a result, the triplet eigenvectors $|\bar{T}_1\rangle$,  
$|\bar{T}_2\rangle$, $|\bar{T}_3\rangle$ can be expressed as Fourier 
transforms  of the basis states $|T_j\rangle$ in the same way the  
single-particle molecular orbitals are expressed in terms of localized  
orbitals $|j\rangle$, shown in Eq.~(\ref{eigenvectors}).  
The two electrons either move clockwise, counterclockwise, or stand 
still.   
The three eigenenergies corresponding to these eigenvectors are,  
respectively,  
$E_{T}^1=2E+V+2|t|\cos\left({2\pi}\phi/3\right)$,  
$E_{T}^2=2E+V+2|t|\cos\left[{2\pi}(\phi+1)/3\right]$,  
and  
$E_{T}^3=2E+V+2|t|\cos\left[{2\pi}(\phi-1)/3\right]$.  
Note the difference in sign of $t$ in the eigenvalues with respect to  
the single electron case.  
As a result, at zero magnetic field we obtain the doubly degenerate 
lowest-energy state $E_T=2E+V-|t|$, and a non-degenerate excited state 
with energy $E_T=2E+V+2|t|$,   
As the magnetic field increases, the triplet energies oscillate with  
the period of $3$ flux quanta.  
  
\begin{figure}  [h]
  \includegraphics[width=0.8\textwidth]{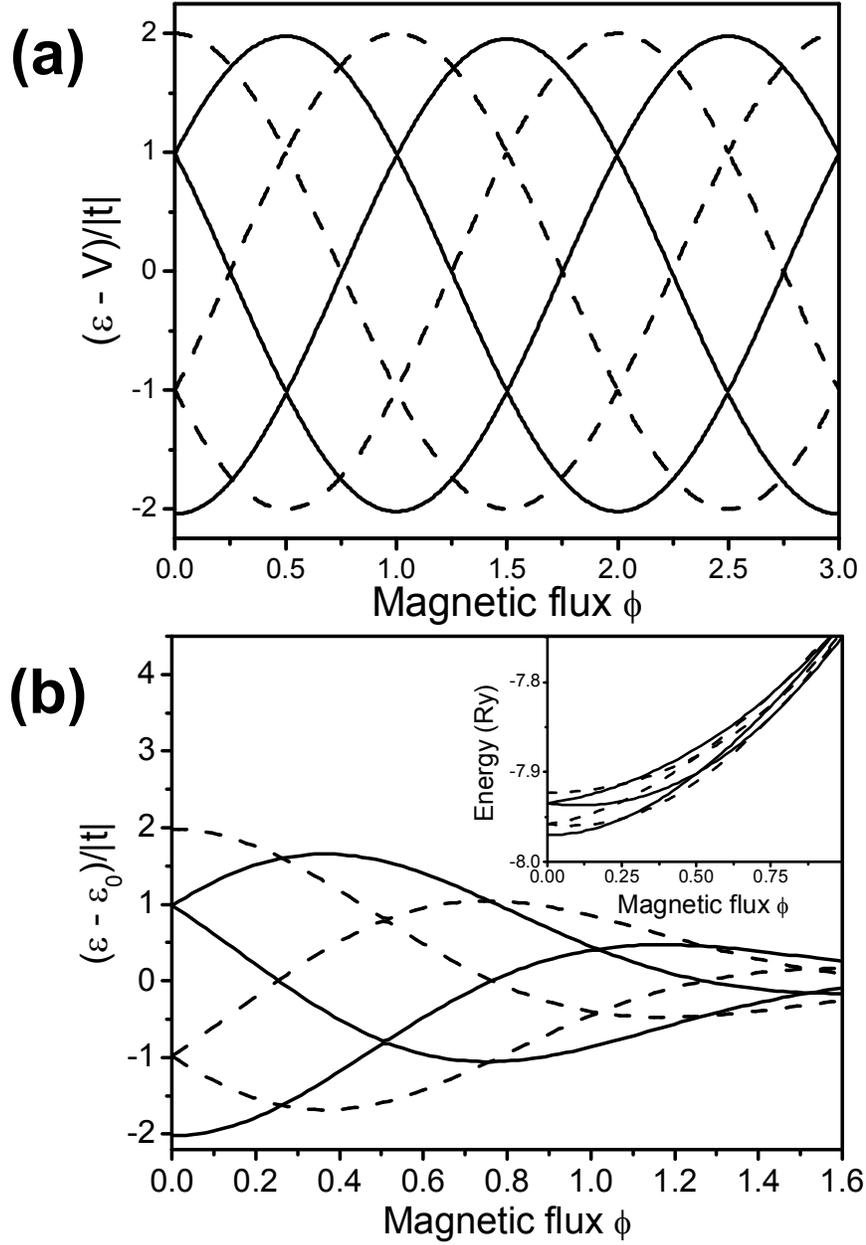}
  \caption{(a) Energy spectrum of the TQD with $N=2$  
    as a function of the number of magnetic flux quanta calculated  
    within the Hubbard model.  
    Solid lines correspond to the triplet levels, while dashed  
    lines correspond to the singlets. (b) The same  
    energy spectrum calculated with the LCHO-CI approximation after  
    removing the diamagnetic shift. Inset corresponds to the actual  
    spectrum with diamagnetic shift.}  
  \label{fig3}  
\end{figure}

Let us now move on to the singlet subspace.  
The singly-occupied singlet configurations $|S_1\rangle$,  
$|S_2\rangle$, and $|S_3\rangle$ are obtained from the triplet  
configurations $|T_1\rangle$, $|T_2\rangle$, and $|T_3\rangle$ by  
flipping the spin of one electron and properly antisymmetrizing the  
configurations.  
For example, the configuration  
$|S_1\rangle={1\over\sqrt{2}}\left(c^\dag_{2\downarrow}c^\dag_{1\uparrow}+  
 c^\dag_{1\downarrow}c^\dag_{2\uparrow}  \right)|0\rangle$.  
In the same way,  
$|S_2\rangle={1\over\sqrt{2}}\left(c^\dag_{3\downarrow}c^\dag_{1\uparrow}+  
 c^\dag_{1\downarrow}c^\dag_{3\uparrow}  \right)|0\rangle$  
and  
$|S_3\rangle={1\over\sqrt{2}}\left(c^\dag_{3\downarrow}c^\dag_{2\uparrow}+  
 c^\dag_{2\downarrow}c^\dag_{3\uparrow}  \right)|0\rangle$.  
In addition to the singly-occupied configurations there are also three  
doubly-occupied configurations $|S_4\rangle$, $|S_5\rangle$, and 
$|S_6\rangle$, such that, e.g.,  
$|S_4\rangle=c^\dag_{1\downarrow}c^\dag_{1\uparrow}|0\rangle$.  
The singly-occupied configurations are characterized by energies  
$2E+V$, and, just as the triplets, they are coupled by tunneling  
matrix elements.  
Unlike in the triplet case, however, these off-diagonal elements do not  
acquire the negative sign.  
On the other hand, the energies of all doubly-occupied configurations  
are $2E+U$, i.e., contain the element describing the Coulomb onsite  
repulsion, making these energies larger than those of the  
singly-occupied configurations.  
The Hubbard Hamiltonian does not mix the configurations $|S_4\rangle$,  
$|S_5\rangle$, and $|S_6\rangle$ with each other, but does mix the  
singly and doubly occupied subspaces.  
Here again it is convenient to Fourier transform the singlet basis set into the  
form $\{|\bar{S_1}\rangle,|\bar{S_2}\rangle,|\bar{S_3}\rangle,  
|S_4\rangle,|S_5\rangle,|S_6\rangle\}$.  
In this basis the full singlet Hamiltonian can be written as two  
$3\times3$ block diagonal matrix coupled through terms that account  
for the interactions between singly and doubly occupied  
configurations:  
\beqa  
\hat{H}_S = \left[  
  \begin{array}{cc}  
  \hat{I}{\bf D}_1 & \hat{C}\\  
  \hat{C}^\dag & \hat{I}{\bf D}_2  
  \end{array}\right],  
\label{hsind}  
\eeqa  
where $\hat{I}$ is the $3\times3$ identity matrix, the vector 
${\bf D}_1^T=\left[  
2E+V-2|t|\cos\left(2\pi\phi/3\right),  
2E+V-2|t|\cos\left(2\pi(\phi+1)/3\right),  
2E+V-2|t|\cos\left(2\pi(\phi-1)/3\right)\right]$, and  
the vector ${\bf D}_2^T=\left[2E+U, 2E+U, 2E+U \right]$.  
The coupling matrix $\hat{C}$ is  
\beqa  
\hat{C}= -{\sqrt{8\over3}}|t|\left[  
  \begin{array}{ccc}  
    \cos\left(\frac{2\pi\phi}{3}\right)  
             & e^{i 2\pi\phi/3} &  
                    e^{-i 2\pi\phi/3}\\  
    {1\over2}e^{-i 2\pi\phi/3}\left(1  
               + e^{i 2\pi\left(2\phi-1\right)/3}\right)&  
    e^{i2\pi\phi/3}\left(1+e^{-4\pi i/3}\right) &  
     e^{-i 2\pi(\phi+1)/3}  
                 \left(1+e^{-2\pi i/3}\right) \\  
     {1\over2}e^{-i2\pi\phi/3}\left(1+e^{i2\pi(2\phi+1)/3}\right)&  
     e^{i2\pi\phi/3}\left(1+e^{4\pi i/3}\right) &  
      e^{-i 2\pi(\phi-1)/3}\left(1+e^{2\pi i/3}\right)\\  
  \end{array}  
\right].  
\nonumber  
%\label{hatc}  
\eeqa  
In Fig.~\ref{fig3}(a) we have plotted the low-energy spectrum for the  
TQD with 2 electrons, assuming the Hubbard parameters $V=0.42$~${\cal 
  R}$ and $U=2.56$~${\cal R}$ 
(Ref.~\onlinecite{korkusinski_gimenez_prb2007}).   
The three dashed lines correspond to the eigenvalues of the triplet  
Hamiltonian while the three solid lines are eigenvalues of the singlet  
Hamiltonian.  
For the $6\times6$ singlet Hamiltonian, there is an additional 
three-fold  degenerate and non-oscillating eigenvalue at higher 
energy, originating from the doubly-occupied configurations (not shown 
in the figure).   
The main result, apparent in  Fig.~\ref{fig3}(a), 
is the existence of transitions between the spin singlet and 
triplet, occurring periodically as a function of the magnetic flux. 
In the region between $\phi=0$ and $\phi=1$ the ground state is a 
singlet for $\phi < 1/4$ and $\phi > 3/4$, and triplet for 
$1/4 < 3\phi/4$. 
This alignment of phases repeats for each subsequent flux quantum. 
 
Upon the inclusion of Zeeman energy we find that the intervals of 
stability of the singlet phase decrease in each subsequent period. 
The spin oscillations are eventually suppressed leading to a 
continuous triplet ground state at sufficiently high magnetic fields.  
  
The existence of spin oscillations is confirmed by results of the 
LCHO-CI calculation presented in Fig.~\ref{fig3}(b).  
The inset of Fig.~\ref{fig3}(b) shows that the diamagnetic shift 
together with the decrease of tunneling between dots makes it 
difficult to distinguish more than one oscillation.  
But after removing the diamagnetic shift by subtracting the reference 
energy $\varepsilon_0$, defined by the ground state energy of the 
two-electron system  without tunneling, the resulting energy spectrum 
[main panel in Fig.~\ref{fig3}(b)] agrees well with the Hubbard model 
except for the exponential decay of the amplitude of energy 
oscillations. 
  
\subsection{Three electrons}  
  
The three electron case at zero magnetic field has been analyzed in detail  
in Ref.~\onlinecite{korkusinski_gimenez_prb2007}.  
Following that scheme, we start our treatment with the completely 
spin-polarized system, i.e., one with total spin $S=3/2$.  
In this case we can distribute the electrons on the three dots in only  
one way: one electron on each site with parallel spin, which gives a 
spin-polarized state 
$|a_{3/2}\rangle=c^\dag_{3\downarrow}c^\dag_{2\downarrow}c^\dag_{1\downarrow}|0\rangle$. 
This is an eigenstate of our system with energy  
$E_{3/2}=3E+3V$.  
Let us now flip the spin of one of the electrons.  
This electron can be placed on any orbital, and with each specific  
placement the remaining two spin-down electrons can be distributed in  
three ways.  
Altogether we can generate nine different configurations.  
Three of these configurations involve single occupancy of the  
orbitals.  
They can be written as  
$|a\rangle=c^\dag_{3\downarrow}c^\dag_{2\downarrow}c^\dag_{1\uparrow}|0\rangle$,  
$|b\rangle=c^\dag_{1\downarrow}c^\dag_{3\downarrow}c^\dag_{2\uparrow}|0\rangle$,  
and  
$|c\rangle=c^\dag_{2\downarrow}c^\dag_{1\downarrow}c^\dag_{3\uparrow}|0\rangle$. 
The remaining six configurations with double occupancy are  
$|d\rangle=c^\dag_{2\downarrow}c^\dag_{1\downarrow}c^\dag_{1\uparrow}|0\rangle$,  
$|e\rangle=c^\dag_{3\downarrow}c^\dag_{1\downarrow}c^\dag_{1\uparrow}|0\rangle$,  
$|f\rangle=c^\dag_{3\downarrow}c^\dag_{2\downarrow}c^\dag_{2\uparrow}|0\rangle$,  
$|g\rangle=c^\dag_{1\downarrow}c^\dag_{2\downarrow}c^\dag_{2\uparrow}|0\rangle$,  
$|h\rangle=c^\dag_{1\downarrow}c^\dag_{3\downarrow}c^\dag_{3\uparrow}|0\rangle$,  
$|j\rangle=c^\dag_{2\downarrow}c^\dag_{3\downarrow}c^\dag_{3\uparrow}|0\rangle$.  
All these configurations are characterized by the same projection of  
total spin, $S_z=-1/2$.  
Moreover, the doubly-occupied configurations are also the eigenstates  
of total spin, with $S=1/2$, while the total spin of the 
singly-occupied configurations is not defined. 
In the basis of the nine configurations we construct the  
Hamiltonian matrix by dividing the $9$ configurations into three groups,  
each containing one of the singly-occupied configurations $|a\rangle$,  
$|b\rangle$, and $|c\rangle$, respectively.  
By labeling each group with the index of the spin-up electron, the  
Hamiltonian takes the form of a $3\times 3$ matrix:  
\begin{equation}  
  \hat{H}_{1/2}=\left[  
    \begin{array}{ccc}  
      \hat{H}_1 & \hat{T}_{12} & \hat{T}_{31}^\dag\\  
      \hat{T}_{12}^\dag & \hat{H}_2 & \hat{T}_{23}\\  
      \hat{T}_{31} & \hat{T}_{23}^\dag & \hat{H}_3\\  
    \end{array}  
    \right].  
    \label{hubbard3e}  
\end{equation}  
The diagonal matrix, e.g.,  
\begin{widetext}  
$$  
\hat{H}_1 = \left[  
  \begin{array}{ccc}  
    3E+2V+U & t_{23} & -t_{13}\\  
    t_{23}^* & 3E+2V+U & t_{12} \\  
    -t_{13}^* & t_{12}^* & 3E+3V \\  
  \end{array}  
\right]  
$$  
\end{widetext}  
describes the interaction of three configurations which contain  
spin-up electron on site 1, i.e., two doubly-occupied configurations  
$|d\rangle$ and $|e\rangle$, and a singly-occupied  
configuration  $|a\rangle$.  
The remaining matrices corresponding to spin-up electrons localized on  
sites 2 and 3 can be constructed in a similar fashion.  
The interaction between them is given in terms of effective and magnetic field dependent hopping  
matrix  
$$  
\hat{T}_{ij} = \left[  
  \begin{array}{ccc}  
    0 & -t_{ij} & 0\\  
    0 & 0 & -t_{ij}\\  
    +t_{ij} & 0 & 0\\  
  \end{array}  
\right].  
$$  
Upon diagonalization of the Hamiltonian (\ref{hubbard3e}) we obtain  
nine levels, of which one corresponds to the total spin $S=3/2$,  
and eight - to the total spin $S=1/2$.  
The energy of the high-spin state is the same as that of the  
configuration $|a_{3/2}\rangle$ discussed above, except for the Zeeman  
contribution, which is different due to the different spin projection  
$S_z$ of the two configurations.  
  
\begin{figure}[h]
  \includegraphics[width=0.8\textwidth]{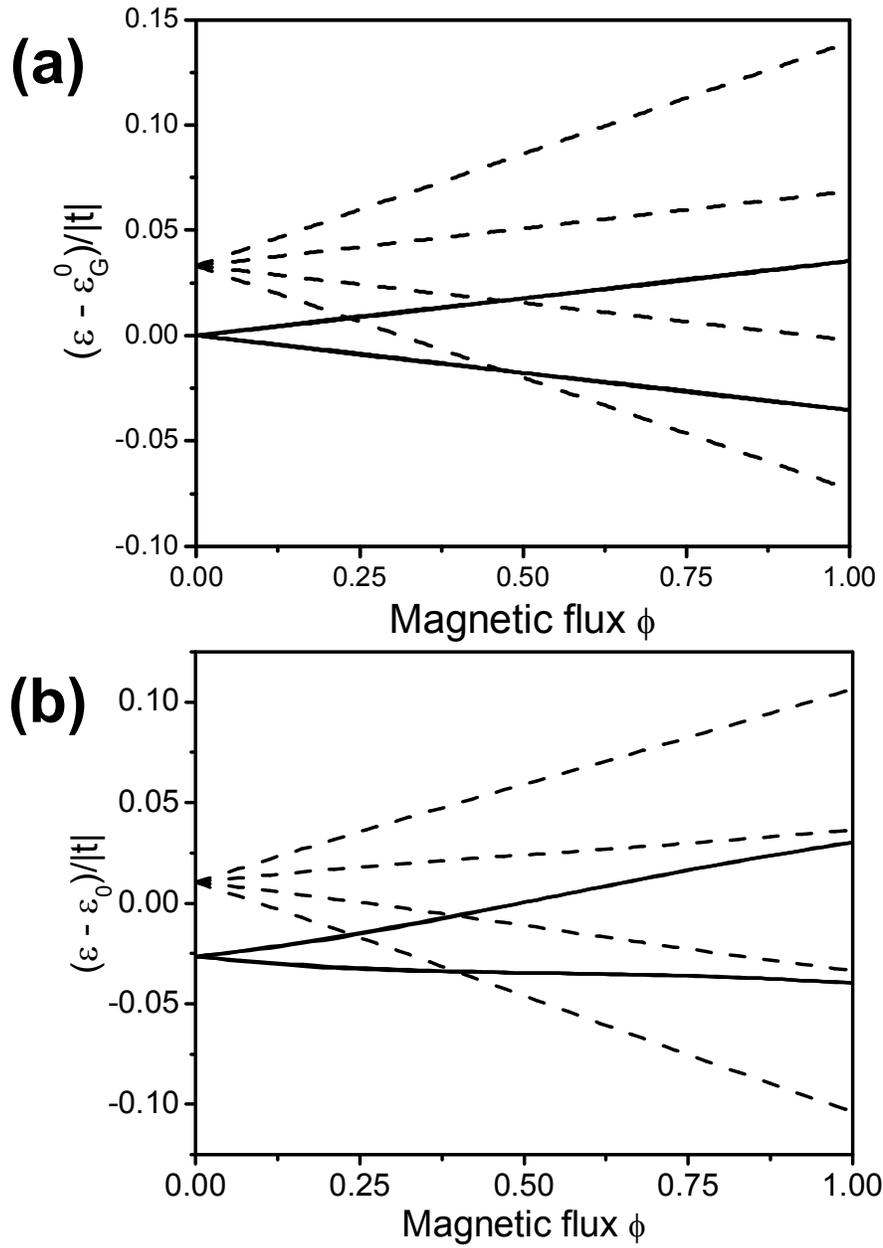}
  \caption{(a) Energy spectrum of the TQD with three electrons  
    versus the dimensionless flux $\phi$ including the Zeeman term.  
     All energies are measured from the ground state energy at $\phi=0$.  
    (b) The same spectrum calculated using the LCHO-CI technique 
    without diamagnetic shift.}  
  \label{fig4}  
\end{figure}

Let us now discuss the energy spectrum of the system at zero magnetic 
field. 
In Ref.~\onlinecite{korkusinski_gimenez_prb2007} we have shown that 
this spectrum is composed of two segments. 
In the low-energy region we find two $S=1/2$, $S_z=-1/2$ states, which 
form a degenerate pair for the TQD on resonance. 
At the energy equal to $3J/2$, where $J$ is the exchange energy, we 
find one $S=3/2$, $S_z=-1/2$ state.  
These three levels are built out of singly-occupied configurations. 
The high-energy part of the spectrum consists of three pairs of 
states, composed of doubly-occupied configurations. 
The two parts of the spectrum are separated by an energy gap 
proportional to the onsite Coulomb element $U$. 
In the following we shall focus on the low-energy segment of the 
spectrum only, shown in Fig.~\ref{fig4}(a). 
 
Before we discuss the three-electron spectrum at finite magnetic 
field, we first account for the correct degeneracy of the energy 
levels by including states with all possible orientations of the total 
spin $S_z$. 
In this case we have two pairs of states with low spin: one pair with 
$S_z=+1/2$, and another with $S_z=-1/2$. 
These two pairs form a degenerate quadruplet at $\phi=0$. 
The high-spin state, on the other hand, is a manifold of four states, 
with $S_z=\pm 3/2$ and $S_z=\pm 1/2$. 
Let us now consider the spectrum at finite magnetic fields accounting 
for the Zeeman energy. 
%As it can be seen in Fig.~\ref{fig4}(a), 
The quadruply degenerate 
low-spin state splits into two branches separated by the Zeeman 
energy, see  Fig.~\ref{fig4}(a),reflecting the different orientations of $S_z$. 
Further, the energies of the states composing each pair oscillate with 
the magnetic field, and cross each other at $\phi=n\pi/2$, 
$n=0,1,\dots$ (not seen on this scale). 
These oscillations have a period $\Delta\phi=1$,  different from the 
period of three flux quanta present for one and two electrons.  
The amplitude is also a non-trivial function of the hopping parameters 
$t$, being more than two orders of magnitude smaller than $|t|$. 
As for the high-spin state, its four-fold degeneracy is lifted by the 
Zeeman energy, but the constituent levels do not exhibit any 
oscillations. 
With increasing magnetic field the $S=3/2, S_z = 3/2$ spin-polarized   
state lowers its energy with respect to the ground  $S=1/2, S_z=1/2$ 
state, and at a critical value of the magnetic field becomes the 
ground state.   
The critical magnetic field $B_c$ is given by the condition $ g^* \mu 
B_c = 3 J /2 $, under which the Zeeman energy equals the exchange 
energy.   
The results of the Hubbard calculations are in agreement with the 
three-electron energy spectra calculated within the LCHO-CI 
approach, shown in Fig.~\ref{fig4}(b). 
Note that a similar analysis was reported in
Ref.~\onlinecite{scarola_dassarma_pra2005} for the TQD composed of
shallower, parabolic dots with larger interdot tunneling.
These calculations revealed an additional total spin oscillation
between the $S=1/2$ and $S=3/2$ phases, which occurs for the same $S_z$
component.
We were able to reproduce this oscillation within the LCHO-CI approach
using shallower dots, but not within the Hubbard model.
This is because the spin oscillation is due to the
magnetically-induced reduction of the interdot tunneling element.
However, the importance of these oscillations is minor due to the
dominant role of the Zeeman energy.
As discussed above, the Zeeman energy leads to the onset of a spin
polarized phase at a sufficiently high magnetic field, which
suppresses any spin oscillations.

\subsection{Four electrons}  
  
The four-electron configurations correspond to two  
holes, created when two electrons are removed from the filled-shell  
configuration.  
With two holes we can form only the spin singlet and triplet  
configurations of the system.  
  
Let us focus on the triplets first.  
They involve one electron spin-up occupying the first, second, or  
third dot in the presence of an inert core of three spin-down electrons.  
If we denote the operator creating a hole on dot $i$ with spin $\sigma$ by  
$h_{i\sigma}^\dag$, we can write the three basis configurations in this  
subspace in the form  
$|T^{(H)}_1\rangle = h^\dag_{1\downarrow}h^\dag_{2\downarrow}|N_e=6\rangle  
=c^\dag_{3\uparrow}c^\dag_{3\downarrow}c^\dag_{2\downarrow}c^\dag_{1\downarrow}|0\rangle$,  
$|T^{(H)}_2\rangle=h^\dag_{3\downarrow}h^\dag_{1\downarrow}|N_e=6\rangle$ and  
$|T^{(H)}_3\rangle=h^\dag_{3\downarrow}h^\dag_{2\downarrow}|N_e=6\rangle$.  
The two-hole triplet Hamiltonian is given by  
\begin{equation}  
\hat{H}_T = \left[  
  \begin{array}{ccc}  
    4E+U+5V & t_{23} & t_{13}\\  
    t_{23}^* & 4E+U+5V & t_{12} \\  
    t_{13}^* & t_{12}^* & 4E+U+5V\\  
  \end{array}  
\right].  
\label{triphamilh}  
\end{equation}  
  
Note that the above Hamiltonian differs from that describing the  
two-electron triplet subspace in that the off-diagonal tunneling matrix  
elements do not acquire the additional negative phase.  
  
Let us move on to the two-hole singlet configurations.  
The singly-occupied states involve the two holes occupying two different  
dots, while the doubly-occupied states hold both holes on the same  
dot.  
This situation is analogous to the two-electron case described  
earlier, and the two-hole singlet Hamiltonian written in the  
appropriately rotated basis is analogous to that shown in  
Eq.~(\ref{hsind}).  
The only difference is that the diagonal vectors ${\bf D}_1$ and  
${\bf D}_2$ will contain two-hole, instead of two-electron energies:  
$4E+U+5V$ instead of $2E+V$ for singly-occupied configurations,  
and $4E+2U+4V$ instead of $2E+U$ for doubly-occupied configurations.  
Also, the factors $|t|$ in ${\bf D}_1$ acquire the opposite sign,  
while this additional phase does not appear in the coupling matrix  
$\hat{C}$ for the two-hole case.  
  
\begin{figure}[h]
  \includegraphics[width=0.8\textwidth]{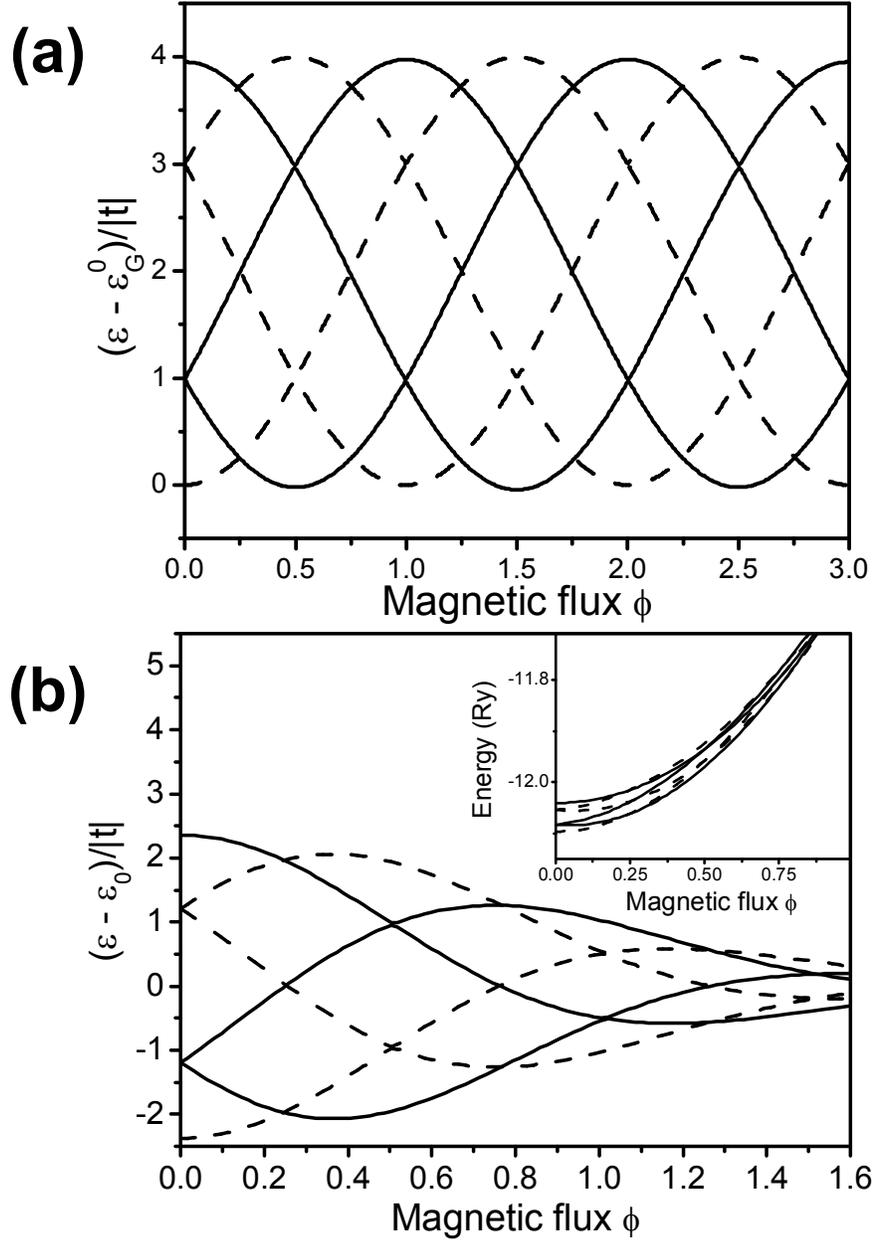}
  \caption{(a) Energy spectrum of the TQD with four electrons  
    versus the dimensionless flux $\phi$ in the absence of Zeeman  
    energy. Triplet levels plotted with dashed lines while singlets  
    are indicated with solid lines.  
    (b) The four-electron spectrum obtained using the LCHO-CI technique  
    with (inset) and without (main panel) diamagnetic shift.}  
  \label{fig4e}  
\end{figure}

In Ref.~\onlinecite{korkusinski_gimenez_prb2007} we have diagonalized  
the singlet and triplet Hamiltonians at zero magnetic field.  
We found that in this case the total spin of the two-hole ground state  
depended on the interplay of Hubbard parameters.  
For a typical case of $2|t|<U-V$ the ground state is a spin triplet.  
Clearly, the appearance of the finite magnetic moment of the ground  
state is made possible by the degeneracy of the single-particle  
excited state at zero magnetic field.  
As we increase the field this degeneracy is removed, so we  
may expect a transition to a spin singlet.  
The periodic reappearance of the degeneracy should lead to spin  
oscillations.  
This is indeed what we observe in the energy spectrum, whose  
low-energy segment is plotted in Fig.~\ref{fig4e}(a) as a function of  
the number of magnetic flux quanta.  
Again, singlet eigenvalues have been plotted with solid lines and dashed  
lines for the triplets.  
As for the case of $N=2$, transitions between triplet and singlet  
ground state appear as we increase the magnetic flux, except 
that the alignment of phases seen for $N=2$ electrons is inverted. 
Furthermore, as in the case of two electrons, the introduction of the  
Zeeman term will favor the spin alignment of the triplet  
configuration, suppressing the triplet-singlet transitions at high  
magnetic field.  
These predictions of the Hubbard model are confirmed by the LCHO-CI  
calculation, whose results are presented in Fig.~\ref{fig4e}(b).  
Again, the original spectrum is shown in the inset, while the main 
panel shows the energies without the diamagnetic shift.  
  
\subsection{Five electrons \label{five}}

Five electrons correspond to a single hole.  
The single-hole Hamiltonian can be obtained from the  
single-electron Hamiltonian by appropriately modifying the diagonal  
terms and setting $t_{ij}\leftrightarrow -t_{ij}$ 
(Ref.~\onlinecite{korkusinski_gimenez_prb2007}).   
For the triangular triple dot on resonance this symmetry is  
reflected in the energy spectrum of the hole as shown in 
Fig.\ref{fig4}(a).  
For the one-hole problem at zero magnetic field,  
the opposite sign of the off-diagonal element leads to a  
doubly-degenerate hole ground state.  
This behavior is confirmed in the LCHO-CI calculations, whose results  
are shown in Fig.~\ref{fig4}(b).  
  
\section{Charging diagram of the resonant triple dot \label{seccharging}}

We can now construct the charging diagram of the triple dot molecule  
as a function of the magnetic field.  
For any number of electrons $N$ (1 to 6) and any quantum dot energy  
$E$, we obtain the ground-state energy $E_{GS}(N)$ and the  
corresponding total spin  by diagonalizing the  
Hubbard Hamiltonian.  
We use these energies to calculate the chemical potential of the  
triple quantum dot molecule $\mu(N)= E_{GS}(N+1)- E_{GS}(N)$.  
When $\mu(N)$ equals the chemical potential $\mu_L$ of the leads,  
the $N+1$st electron is added to the $N$-electron quantum-dot  
molecule.  
This establishes the total number of electrons $N$ in the quantum  
dot molecule and their total spin as a function of  
quantum-dot energy $E$ relative to the chemical potential of the leads 
$\mu_L$.   
In the case of LCHO-CI approach, the relevant quantum dot energy  
is the single-particle reference energy $E$.  
Changes in electron numbers can be detected by Coulomb  
blockade (CB), spin blockade, or charging  
spectroscopies.\cite{pioro_abolfath_prb2005,gaudreau_studenikin_prl2006}  
The calculated stability diagram, with Hubbard parameters used in the  
previous section and taking into account the contribution of the  
Zeeman term, is shown in Fig.~\ref{fig5}(a), while Fig.~\ref{fig5}(b)  
shows the stability diagram computed using the LCHO-CI approach.  
Note that in this figure the oscillations of the stability lines  
corresponding to the condition $\mu(N)-\mu_L=0$ are not visible due to  
the energy scale.  
As explained in the previous sections, the differences among both 
addition spectra are due to the diamagnetic shift and the  
suppression of the inter-dot tunneling. 
These effects appear naturally in the LCHO-CI approach but are not 
taken into account in the Hubbard model.  
  
Let us explain the addition spectrum as we change the single-dot 
energy $E$ of each dot with respect to the chemical potential of the  
leads $\mu_L=0$.  
From the condition $\mu_L = E-2|t|$, the energy $E_{GS}(1)$  
corresponding to the addition of the  
first electron can be approximated by $E(1)\approx 2|t|$.  
At this energy the first Coulomb blockade peak of the triple quantum  
dot molecule should be observed.  
Similar arguments based on the results of the previous section  
can be applied to find the CB peaks corresponding to the addition  
of the remaining electrons.  
Note that the prominent energy gap that appears for the addition of 
the fourth electron on the TQD is due to the large on-site Coulomb 
repulsion $U$.   
The ground state of the three-electron TQD corresponds to one electron 
occupying each dot, and therefore the addition of a new electron will  
increase the energy of the system by order of $U$.  
This term does not appear in any other addition processes,  
in which only the interdot Coulomb element $V$ is relevant. 
  
The spin oscillations in the system with two and four electrons, as 
well as the spin transition in the three electron system,  
lead to a strong modulation of the current through the TQD.  
An schematic representation of the amplitude of Coulomb blockade peaks 
as a function of the magnetic flux is given in Fig.~\ref{fig6}, for 
different numbers of electrons confined in the triple dot.  
Here we assume that the leads are spin-unpolarized, and the transport 
involves only the lowest-energy level of the molecule. 
The vertical arrow shows the transition in the three electron system 
from $S=1/2$ to $S=3/2$ state.   
When the two  electrons are in a spin triplet with $S=1$, adding an 
electron can create a spin polarized final $S=3/2$ three electron 
droplet, and the current is high.  
However, if the two-electron system is in a singlet $S=0$ state, the 
final $S=3/2$ state cannot be reached by adding a single electron, and 
the current is spin blockaded.  
Hence quenching of the tunneling current in the two-electron droplet 
is a signature of a spin-polarized three-electron state   
and a spin-singlet two-electron state.  
Adding a fourth electron to a three-electron droplet is equivalent to 
adding a hole to a two-hole droplet.  
Hence the oscillation in CB peak amplitude, but shifted in phase  
since the holes start as triplets and electrons start as singlets.   
  
\section{Conclusions\label{summary}}  
  
In conclusion, we have studied the effect of the magnetic field on the 
electronic properties of a triple triangular quantum dot molecule.  
Exact results for few-electron spectra in a magnetic field where 
obtained  for identical dots in the Hubbard model.  
Aharonov-Bohm oscillations of a single electron, singlet-triplet spin 
oscillations for pairs of electrons and pairs of holes,   
as well as a transition from the frustrated magnetic state to spin 
polarized state for a half-filled lowest electronic shell are 
predicted.  
The impact of spin transitions on the stability diagram and modulation 
of the current through the TQD molecule with increasing magnetic flux 
are discussed.   
The results of the Hubbard model are supported by a microscopic 
calculation using  a general LCHO-CI approach extended to finite 
magnetic fields.   
  
\section*{Acknowledgments}  
The Authors thank A. Sachrajda, L. Gaudreau, and S. Studenikin for 
discussions. 
P.H. and Y.-P.S. acknowledge support by the Canadian Institute for 
Advanced Research.   
F.D. acknowledges partial financial support from Ministerio de  
Educaci\'on y Ciencia, Spain, under grant No. EX2006-0587.

\newpage

\begin{figure}[h]
  \includegraphics[width=0.8\textwidth]{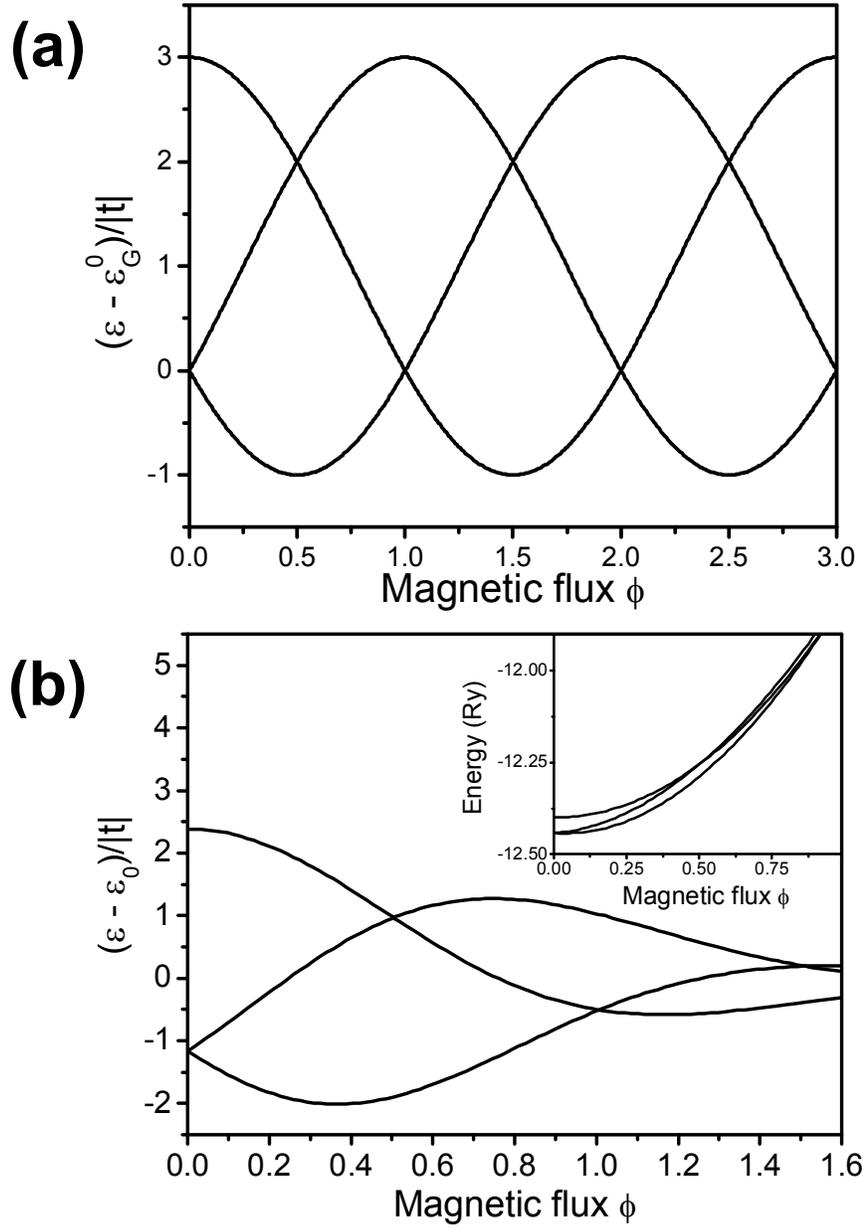}
  \caption{(a) Energy spectrum of the TQD with five electrons  
    versus the dimensionless flux $\phi$, plotted with the zero-field  
    ground-state energy treated as reference.  
    (b) The five-electron spectrum obtained using the LCHO-CI method  
    with (inset) and without (main panel) diamagnetic shift.}  
  \label{fig2}  
\end{figure}  
  
\begin{figure}[h]
  \includegraphics[width=0.8\textwidth]{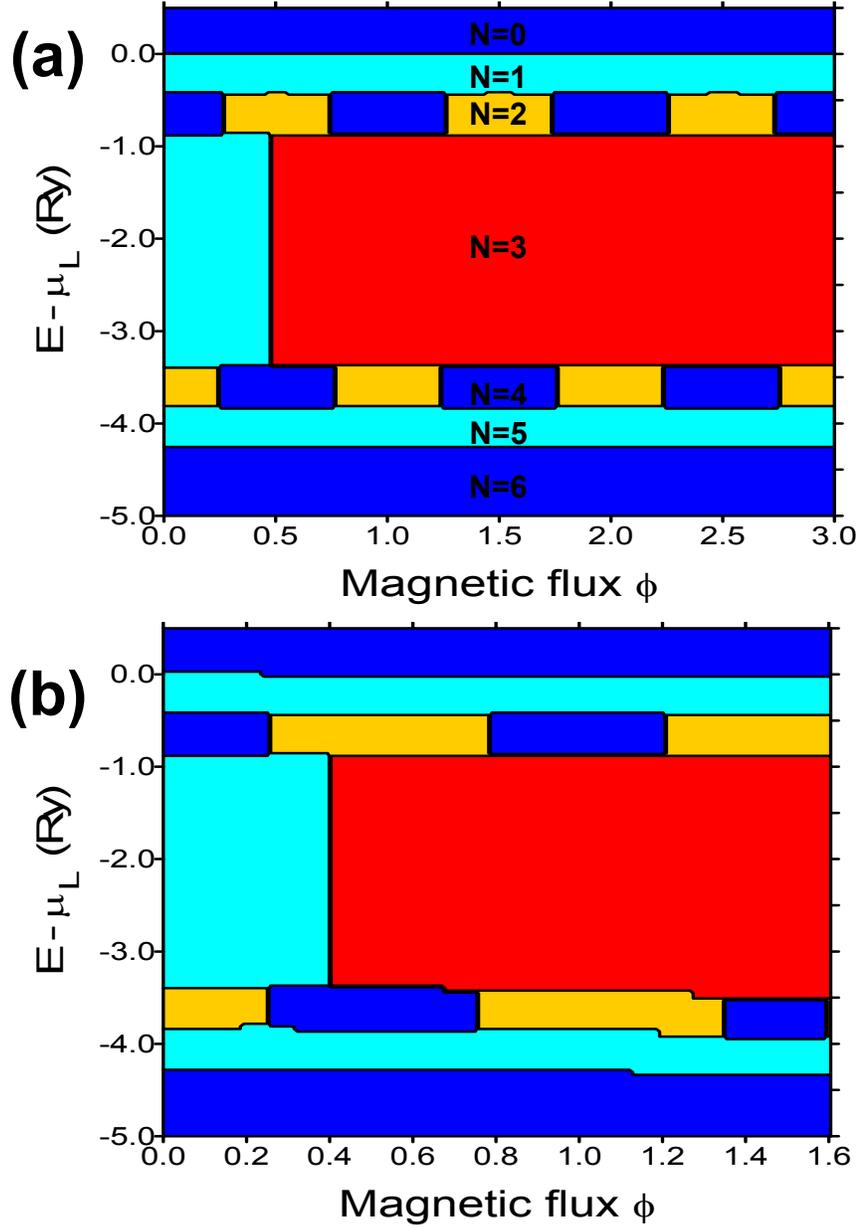}
  \caption{(Color online) (a) Charging diagram for the TQD  versus the 
    number of magnetic flux quanta $\phi$ calculated in the Hubbard 
    model.   
    Colors indicate the total spin of the ground state:  
    $S=0$ (blue), $S=1/2$ (cyan), $S=1$ (yellow), and $S=3/2$ (red).  
    (b) The diagram obtained using the LCHO-CI method.}  
  \label{fig5}  
\end{figure}

\begin{figure}[h]
  \includegraphics[width=0.8\textwidth]{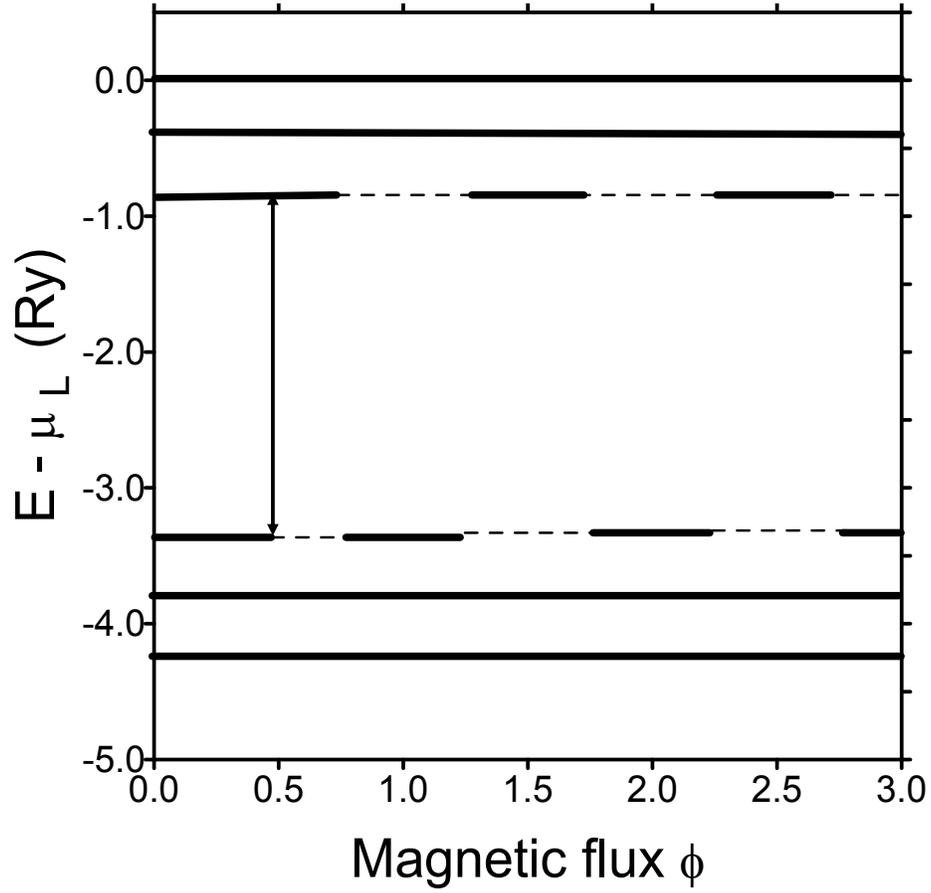}
  \caption{Qualitative scheme of the current passing through  
    the TQD filled with $N$ electrons extracted from the Hubbard results.  
    Thick solid lines correspond to a large tunneling current while 
    thin-dotted lines correspond to a small current under the spin 
    blockade condition.}  
  \label{fig6}  
\end{figure}

\end{document}